\documentclass[twocolumn,showpacs,preprintnumbers,prb]{revtex4-1}

\usepackage{eurosym}
\usepackage{graphicx,bm,amsmath,amssymb}
\usepackage{xcolor}

\def\gz{\ifmmode{Z\hskip -4.8pt Z}
    \else{\hbox{$Z\hskip -4.8pt Z$}}\fi}

\newcommand{\be}{\begin{equation}}
\newcommand{\ee}{\end{equation}}
\newcommand{\bea}{\begin{eqnarray}}
\newcommand{\eea}{\end{eqnarray}}

\begin{document}

\title{Charge and heat pumping in the Rice-Mele chain at finite temperature}

\author{P. Roura-Bas}
\affiliation{Centro At\'{o}mico Bariloche, GAIDI, CONICET,  8400 Bariloche, Argentina}

\author{A. A. Aligia}
\affiliation{Instituto de Nanociencia y Nanotecnolog\'{\i}a 
CNEA-CONICET, GAIDI,
Centro At\'{o}mico Bariloche, 8400 Bariloche, Argentina}

\begin{abstract}
It is well known that quantized topological charge pumping takes place in
the half filled Rice-Mele chain performing a closed cycle in parameter space.
We extend previous studies to the case of charge and heat transport
at arbitrary filling and temperature using the corresponding continuity equation
with focus in the non-interacting case.
The amount of charge and heat transported for any adiabatic time dependence of
the parameters is given by a double integral of an analytical function. We
find that quantized transport is lost except in trivial cases. In
particular, for popular pumping circuits used which lead to quantized non-trivial charge transport at zero temperature, the heat transported
in the cycle vanishes. For other pumping circuits, there is a heat transport
among even and odd sites of the chain and the environment.
As the temperature is increased,
the transported charge and heat decrease and vanish at infinite temperature.
\end{abstract}

\maketitle

\section{Introduction}

\label{intro}

In recent years, topological quantum states of matter have garnered
significant attention due to their fundamentally new physical phenomena and
potential applications in innovative devices \cite{Ando13,Bradlyn17,Sato17}.
An essential concept in the discussion of topological phases is the Berry
phase, a geometric phase acquired by a quantum system during an adiabatic
cycle \cite{Berry84}. Thouless showed that in a periodic system the
geometric phase contributions after one adiabatic cycle are quantized and
correspond to the singularities enclosed by the pump trajectory and the
transported charge \cite{Thou83}. In general, topological quantized charge
or spin pumping can be realized in time-dependent adiabatic evolution in a
closed cycle in a certain space of parameters \cite{Thou83,Niu84,Wang13,Citro23}.

Many topological invariants are extensions of the Berry phase originally
calculated by Zak for a single-particle state as the wave vector traverses
the entire Brillouin zone in one dimension \cite{Zak89} (see for example,
Refs. \cite{Ando13,Tewari12,Asb16,Cardano17,Daroca21,Osta23,Aligia23}). This Berry phase, forms
the foundation of the modern theory of polarization 
\cite{Resta94,Xiao10,Vander18,Bradlyn22} and has been extended to the many-body
context \cite{Ortiz94,resor,oc,Song21}. Changes in polarization are
proportional to the corresponding changes in the Berry phase 
\cite{Ortiz94,Wata18,Aligia23}.

Typically, a two-dimensional
pump cycle encloses one or more singularities where a
symmetry-protected topological number (or geometric phase) undergoes a jump.
Outside these critical points, the protecting symmetry is broken, allowing
for a continuous variation of charge or spin Berry phases. The change in
these phases over the cycle determines the amount of charge or spin
transported. Experimentally, charge transport in Thouless pumps, as
described by the Rice-Mele chain (RMC) \cite{Naka16,Lohse16}, including
cases with interactions \cite{Walter23,Viebahn23}, has been demonstrated in
ultracold atom chains. Theoretically, several variants of the model have also
been explored 
\cite{Asb16,Hay18,Nakag18,Bertok22,Citro23,Roura23,Moreno23,Hattori23,Aligia23,Arg24,Tada24}.

Quantum spin pumps have been also realized experimentally \cite{Schw16} and
discussed theoretically 
\cite{Schw16,Shin5,Meid11,Zhou14,Chen20,Aligia23,Farre24,Marquez24}.

In contrast to charge and spin pumping, heat pumping has not received much
attention and its relation to topological numbers is unclear. However,
beyond its intrinsic interest, heating of the system can be detrimental for
charge or spin pumping. Therefore the study of heat pumping is interesting
from several points of view. 

Recently Hattori \textit{\ et al.} analyzed
charge and energy pumping in the RMC at
finite temperatures \cite{Hattori23}. 
The authors obtained that the pumped charge in a cycle is quantized at
sufficiently low temperature, and find that it vanishes at high temperature.
Instead, they find that the pumped energy is independent of temperature and
explained by the sum of Berry phases of both bands.
Their approach is based in defining macroscopic
charge and energy polarizations proportional to the position 
in space. This approach is at least dangerous and controversial.
For example, a charge distribution on the
boundary of the finite system can lead to a contribution
to the polarization per unit volume which does not
vanish in the thermodynamic limit. 
Therefore, in general the \textit{change} in charge polarization
due to some perturbation (a phonon mode, ferroelectric displacement, etc)
is considered in the modern theory of polarization 
\cite{Asb16,Resta94,Xiao10,Vander18,Bradlyn22,Ortiz94,resor,oc,Song21,Wata18,Aligia23}.
A detail discussion 
is contained in Ref. \onlinecite{Ortiz94}.

In this paper, we derive the heat current traversing a link of the
interacting RMC using the continuity equation. In the non-interacting case,
this current is given by an integral involving the on-site energy and the Berry curvature, for which we
provide an analytical expression for any pumping circuit. While our main
conclusions regarding charge transport at finite temperatures, coincide with
those of Hattori \textit{\ et al.}, we find that the transported heat is in
general not directly related to Berry phases and it depends on
temperature. For simple elliptical circuits used before, the pumped heat is
identically zero. We prove that for any general circuit, the transported
charge, energy and heat are zero for infinite temperature or
completely filled systems.

The paper is organized as follows. In Sec. \ref{model} we present the
Hamiltonian of the model and derive the operators for the different currents.
In Section \ref{geom} we relate these currents with geometrical
concepts such as the Berry curvature and the (charge) Berry phase.
Section \ref{cberry} contains a contour plot of the Berry phase,
which permits to infer the transported charge, energy and heat
for any adiabatic evolution of the parameters of the RMC.
In Section \ref{calc} we present explicit calculations of the transported
quantities for several pumping circuits.
At the end, a summary and discussion can be
found in Sec. \ref{sum}.

\section{Model and current operators}

\label{model}

The Hamiltonian of the interacting RMC can be written in compact form as

\begin{eqnarray}
H &=&\sum\limits_{j=1}^{M}\left[ -t_{0}+\delta \;(-1)^{j}\right]
\sum\limits_{\sigma ={\uparrow ,\downarrow }}\left( c_{j+1\sigma }^{\dagger
}c_{j\sigma }+\text{H.c.}\right)   \notag \\
&&+\Delta \sum\limits_{j=1}^{M}
\sum\limits_{\sigma ={\uparrow ,\downarrow }}(-1)^{j}n_{j\sigma }+U\sum\limits_{j=1}^{M}n_{j\uparrow }n_{j\downarrow },
\label{hirm}
\end{eqnarray}
where $n_{j\sigma }=c_{j\sigma }^{\dagger }c_{j\sigma }$ is the operator of
the number of electrons with spin $\sigma $. An alternative notation for the
hopping terms that we shall use below is \cite{Asb16}
\begin{equation}
v=-t_{0}-\delta ,\text{ }w=-t_{0}+\delta .  \label{uv}
\end{equation}
In an adiabatic evolution, all parameters can depend on time $t$.

The number of sites $M$ is even and the system consists of $M/2$ unit cells
labeled by $m$. The $m$th unit cell contains two sites with $j=2m-1$ and 
$j=2m$.
For the sake of clarity in the following argument, we assume for the moment
a chain with open boundary conditions. Clearly the charge and energy
currents in the middle of the chain do not depend on the boundary conditions
in the thermodynamic limit.
Following Ref. \onlinecite{Asb16}, the particle (charge) current between
the cells $m$ and $m+1$ with spin $\sigma $ can be calculated from the
continuity equation

\begin{equation}
J_{p\sigma }=-i[\hat{N}_{m},H(t)],\text{ }
\hat{N}_{m}=\sum\limits_{j=1}^{2m} n_{j\sigma },  \label{jp}
\end{equation}
where $\hat{N}_{m}$ is the total number of particles at the left of the link
between cells $m$ and $m+1$, and we take $\hslash =1$. Evaluating the
commutator, the operator becomes

\begin{equation}
J_{p\sigma }=-iw(t)\left( c_{2m+1\sigma }^{\dagger }c_{2m\sigma }
-\text{H.c.}\right) .  \label{jp2}
\end{equation}

In a similar way, the operator for the energy current for spin $\sigma $
can be evaluated from the time dependence of $H_{m}$, which is given by 
Eq. (\ref{hirm}) replacing $N$ by $2m$. The result is

\begin{equation}
J_{E\sigma }=\left[ U(t)n_{2m\bar{\sigma}}-\Delta (t)\right] J_{p\sigma },
\label{je}
\end{equation}
where $\bar{\sigma}$ is the spin opposite to $\sigma $. The heat current is 
$J_{Q\sigma }=J_{E\sigma }-\mu J_{p\sigma }$, where $\mu $ is the chemical
potential. For a half filled system $\mu =U/2$. 

For a finite open chain, the above currents depend on $m$. However, in the
thermodynamic limit the result should be independent of $m$ unless the link
is near to the ends. In fact the properties of the system for large $N$
should be independent of the boundary conditions and for periodic boundary
conditions (PBC) the result is independent of $m$. However, a different
result is expected for the current between two sites inside a unit cell 
($j=2m-1$ and $j=2m$). Because of the symmetry of the Hamiltonian, the
expectation value of the corresponding currents 
$\tilde{J}_{p\sigma },\tilde{J}_{E\sigma }$ 
for PBC are given by those of $J_{p\sigma },J_{E\sigma }$
interchanging $v\leftrightarrow w$ and changing the sign of $\Delta$.

A note on the conservation of particles and energy is in order here.
One can apply the same reasoning as above, but starting from the occupancy and energy on the right side of the link (rather than the left). 
As expected,
the change in the number of particles at the right of the link
is opposite to that at the left of the link, confirming the conservation of particle number. 
However, the changes in energy behave differently.
In particular, for $U=0$ they are are identical on both sides of the link. 
In fact, transferring an electron from a site with on-site
energy $-\Delta$ at the left of the link to a site with energy $\Delta$ at 
the right, increases the energy of both sides by the same amount $\Delta$.
It is important to note that to alter the parameters of the model with time, the system must exchange energy with its external environment. For example if $\Delta$ decreases
as a consequence of changes in an optical lattice with cold atoms \cite{Walter23,Viebahn23}, the ground-state energy of the Hamiltonian Eq. (\ref{hirm})
increases and therefore energy should be taken from the laser beam.

In the following we use PBC and discuss mainly the non-interacting ($U=0$) case at half filling, for which $\mu =0$ and therefore $J_{Q\sigma }=J_{E\sigma }$.

\section{Relation with geometric properties }

\label{geom}

\bigskip In the non-interacting case, the Hamiltonian reduces to a $2 \times 2$
matrix for each wave vector $k$ of the form

\begin{eqnarray}
H_{k} &=&d_{x}(k)\sigma _{x}+d_{y}(k)\sigma _{y}+d_{z}(k)\sigma _{z},  \notag
\\
d_{x} &=&v+w\cos (k),\text{ }d_{y}=w\sin (k),  \notag \\
d_{z} &=&\Delta ,  \label{hk}
\end{eqnarray}
where $\sigma _{\alpha }$ are the Pauli matrices.

For the sake of clarity, we begin reviewing the case of a half-filled system
at zero temperature. It has been shown (section 5.3 of Ref. \onlinecite{Asb16})
that the particle current in this case, is given by the integral of the
Berry curvature. Since for $U=0$, the spin is irrelevant, we drop the
subscript $\sigma $ for simplicity. We also denote the charge Berry phase
simply as Berry phase. The static solution of the matrix has a ground state
$|u_{1}(k,t)\rangle $ and an excited state $|u_{2}(k,t)\rangle $. During the
time evolution, the actual ground state $|\tilde{u}_{1}(k,t)\rangle $ is
modified by a mixture with the excited state $|u_{2}(k,t)\rangle $. For
sufficiently slow evolution one has

\begin{eqnarray}
|\tilde{u}_{1}\rangle  &=&e^{-i\varphi }\left[ |u_{1}\rangle +i\frac{\langle
u_{2}|\partial _{t}|u_{1}\rangle }{E_{2}-E_{1}} |u_{2}\rangle \right],  \notag \\
\varphi (t) &=&\int_{0}^{t}dt^{\prime }E_{1}(k,t^{\prime }),  \label{utilde}
\end{eqnarray}
where $E_{n}$ is the energy of the band $n$ for wave vector $k$ at time $t$
given by%
\begin{eqnarray}
E_{1(2)}(k,t) &=&-(+)[\Delta ^{2}+(v+w\cos (k))^{2}  \notag \\
&&+(w\sin (k))^{2}]^{1/2},  \label{ene}
\end{eqnarray}
This leads to a non trivial expectation value of the current for the half
filled case

\begin{equation}
\left\langle J_{p}(t)\right\rangle =\frac{1}{2\pi }\int dkj(k,t),
\text{ }j(k,t)=\langle \tilde{u}_{1}(k,t)|J_{p}(t)|\tilde{u}_{1}(k,t)\rangle .
\label{mjp}
\end{equation}
Explicit calculations show that  $j(k,t)$ is given by the Berry curvature of
the lower band $\Omega _{1}(k,t)$ \cite{Asb16}

\begin{equation}
j(k,t)=\Omega _{1}(k,t)=-i\left( \langle \partial _{k}u_{1}|\partial
_{t}u_{1}\rangle -\langle \partial _{t}u_{1}|\partial _{k}u_{1}\rangle
\right) .\text{ }  \label{ome}
\end{equation}
From the general expression for the Berry curvature for two-dimensional
two-band models \cite{Kordon24},
\begin{equation}
\Omega _{1}(x)=\frac{-1}{2\lvert d(x)\rvert ^{3}}\sum_{ijk}\epsilon
_{ijk}d_{i}(x)\frac{\partial d_{j}(x)}
{\partial x_{1}}\frac{\partial d_{k}(x)}{\partial x_{2}},  \label{ome2}
\end{equation}
we derive a general formula for the pump cycle based on the Berry curvature in the RMC,
\begin{equation}
\Omega _{1}=\frac{w\left[ (\dot{v}\Delta - v\dot{\Delta})\cos (k)
+ \dot{w}\Delta - w\dot{\Delta} \right] }{2\left[ v^{2}+w^{2}+\Delta ^{2}+2vw\cos (k)\right]
^{3/2}.},  \label{ome3}
\end{equation}
where the dots on the parameters denote time derivatives.

The total number of particles (charge)  transported between times 0 and
$t$ is the integral of the current $\left\langle J_{p}(t)\right\rangle $ and
using Eqs. (\ref{mjp}) and (\ref{ome}) one obtains

\begin{equation}
\Delta N(t)=\frac{1}{2\pi }\int_{0}^{t}dt^\prime\int_{0}^{2\pi }dk\Omega _{1}(k,t^\prime).
\label{dq}
\end{equation}
Using Eq. (\ref{ome}) in the form $\Omega _{1}(k,t)=-i(\partial _{k}\langle
u_{1}|\partial _{t}u_{1}\rangle -\partial _{t}\langle u_{1}|\partial
_{k}u_{1}\rangle )$ one notes that the integral over $k$ of the first term
vanishes because $\langle u_{1}|\partial _{t}u_{1}\rangle $ is the same for 
$k=0$ and  $k=2\pi $. The integral over $k$ of the second term is just the
time derivative of the Zak Berry phase

\begin{equation}
\gamma =i\int_{0}^{2\pi } dk \langle u_{1}|\partial _{k}u_{1}\rangle .
\label{gam}
\end{equation}
Therefore the number of particles transported is equal to the change in the
Berry phase

\begin{equation}
\Delta N(t)=\frac{1}{2\pi }\int_{0}^{t} dt^\prime \frac{\partial \gamma (t^\prime)}{\partial t^\prime}.  \label{dq2}
\end{equation}
From the modern theory of polarization, this result is known to be valid
quite generally, for any system and in presence of interactions 
\cite{Ortiz94,resor,oc,Song21,Wata18,Aligia23}.  
For a cyclic evolution, after a
period $T$ in which the system returns to the initial state, $\gamma $
should change in $2\pi q$ with $q$ integer, and the transported charge
$\Delta N(T)=q$ is quantized.

We now turn to discuss particle transport at arbitrary temperatures and
chemical potential in the non-interacting RMC. 
The limiting cases are slow thermalization and fast thermalization. 
In the latter case, the occupancy of each state
$f_{n}(k,t)$, with $n \in \ {1,2}$, is given by the Fermi-Dirac distribution 
at each time
\begin{equation}
f_{n}(k,t)=[\text{exp}(E_{n}(k,t)/k_{B}\tau )+1]^{-1}  \label{fi}
\end{equation}
where $E_{n}$ are given by Eq. (\ref{ene}). For slow thermalization,
$f_{n}(k,t)$ is replaced by the distribution
$f_{n}(k,0)$ at the initial time $t=0$ \cite{Wang13}. 
We assume fast thermalization here, using Eq. (\ref{fi}), but the 
results are only weakly dependent on this choice (compare Figs. \ref{2e} top and \ref{slow}).

Clearly, the  transported charge at finite temperature is a simple
generalization of Eq. (\ref{dq})

\begin{equation}
\Delta N(t)=\frac{1}{2\pi }\int_{0}^{t}dt^\prime\int_{0}^{2\pi
}dk \sum\limits_{n}f_{n}(k,t^\prime)\Omega _{n}(k,t^\prime).  \label{dqt}
\end{equation}
$\Omega _{2}(k,t)$ is the Berry curvature of the upper band and is simply
given by

\begin{equation}
\Omega _{2}(k,t)=-\Omega _{1}(k,t)  \label{omex}
\end{equation}
This result can be understood in two straightforward ways. (i) from differential geometry, it is known that the sum of all band curvatures corresponds to the total vector bundle, which in this case is isomorphic to $B\times {\mathbb C}^2$, where $B$ is the manifold with coordinates $(k,t)$.
This bundle is flat, meaning it has vanishing curvature.
ii) one band
can be mapped onto the other by an electron-hole symmetry $c_{j\sigma
}^{\dagger }\leftrightarrow c_{j\sigma }$. The effect of this transformation
in the Hamiltonian Eq. (\ref{hirm}) in the non-interacting ($U=0$) case is
to change the sign of all parameters and this leads to a change of sing in  
$\Omega _{1}$ [see Eq. (\ref{ome3})].

Eqs. (\ref{dqt}), (\ref{ome3}) and (\ref{omex}) provide and expression for
the charge transport in the RMC at arbitrary temperatures and filling in
terms of a double integral of an analytical function, for any smooth
(derivable) variation of the parameters. 

Using Eq. (\ref{je}) for $U=0$, the extension of the above result to the
energy and heat transport is straightforward. In the non-interacting case
(or the interacting one in the mean-field approximation), the energy and
charge current differ in a constant, but the operator part is the same.
Therefore following a similar procedure as above, one obtains for the
transported energy

\begin{equation}
\Delta E(t)=\frac{-1}{2\pi }\int_{0}^{t}\Delta (t^\prime)dt^\prime\int_{0}^{2\pi
}dk \sum\limits_{n}f_{n}(k)\Omega _{n}(k,t^\prime),  \label{det}
\end{equation}
and at zero temperature and half filling

\begin{equation}
\Delta E(t)=\Delta Q(t)=\frac{-1}{2\pi }\int_{0}^{t}dt^\prime\Delta (t^\prime)
\frac{\partial \gamma (t^\prime)}{\partial t^\prime}.  \label{de}
\end{equation}
The presence of $\Delta (t)$ in the integrand impedes to relate directly the
transported heat $\Delta Q(t)$ to a topological integer.

A consequence of Eq. (\ref{omex}) is that for completely filled bands 
($f_{n}(k)=1$) or infinite temperatures ($f_{n}(k)=1/2$), one has
$\Delta N(t)=\Delta E(t)=0$.
In fact, the first result is valid in the general case, including
interactions, because the expectation values of the currents
Eqs. (\ref{jp2}) and (\ref{je}) vanish. Alternatively, the electron-hole transformation maps the completely filled system in the completely
empty one, for which the currents trivially vanish.

\section{Contour plot of the Berry phase}

\label{cberry}

\begin{figure}[bh]
    \centering
    \includegraphics[width=1.0\linewidth]{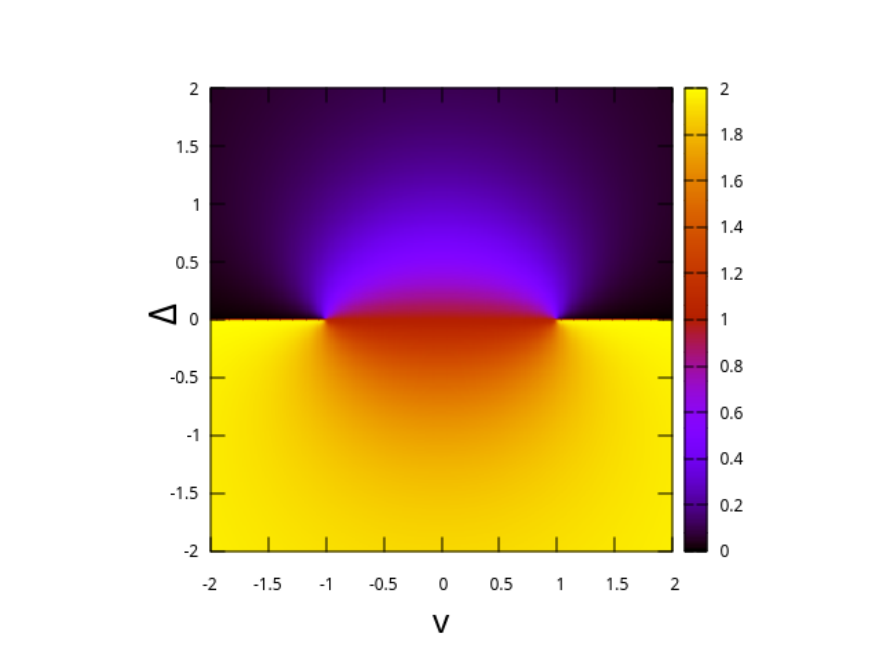}
    \caption{Berry phase $\gamma/\pi$ of the lower band
    as a function of $v$ and
    $\Delta$ for $w=1$ and $U=0$.}
    \label{berry}
\end{figure}

In Fig. \ref{berry} we show the Berry phase $\gamma$ in the
$(v,\Delta)$ plane
fixing the other parameters. Due to Eq. (\ref{dq2}), the number of
particles transported at zero temperate and half filling, for any time evolution of $v$ and $\Delta$,
can be read from this plot. Using Eq. (\ref{de}), the transported
energy and heat can also be inferred from the figure, although
the presence of $\Delta(t)$ inside the integral affects the topological
character of the transport for a closed cycle, as discussed below.

For any closed circuit,
the number of transported particles is quantized reflecting the topological nature of the Thouless pump. We explain this fact
in more detail below.
For $\Delta=0$, the system has inversion symmetry around any point between two atoms. As a consequence, $\gamma$ can only be 0 or
$\pi$ mod $2 \pi$ \cite{Zak89,Aligia23}. Thus $\gamma$ becomes a $Z_2$
topological number. As reflected in Fig. \ref{berry}, $\gamma$ jumps
at two critical points $(v,\Delta)=(\pm 1,0)$, with $\gamma=0$
for $|v|>1$ and $\gamma=\pi$ for $-1<v<1$ mod $2 \pi$ for $\Delta=0$.

For finite $\Delta$ the inversion symmetry is lost and the Berry phase
varies continuously as the parameters are changed. However for any closed circuit in which initial and end points $(v,\Delta)$ coincide,
it is clear that the total number of particles transported
in  a closed circuit $\Delta N_c$ is quantized. Denoting as $l_+$ [$l_-$] the difference between the number of times a closed circuit encloses the point $(1,0)$ [$(-1,0)$]
in the anticlockwise or clockwise direction, using Eq. (\ref{dq2}) one has
\begin{equation}
\Delta N_c=l_+ - l_-
\label{dqb}
\end{equation}

For example, if the circuit is a large circle enclosing both points in the clockwise
direction, it is clear from Fig. \ref{berry} that the Berry phase $\gamma \sim 0$
in the whole circuit and then $\Delta N_c=0$, in agreement with Eq. (\ref{dqb})
with $l_+ = l_-$. If however, one takes a small circle in the anticlockwise direction
surrounding the point (-1,0) and starting at (-0.9,0). The initial value of $\gamma$ is $\pi$, then $\gamma$ decreases being $\sim \pi/2$ at (-1,0.1) and vanishes [mod($ 2 \pi$)] at the point (-1.1,0) where half of the cycle was completed. In the other half
of the cycle $\gamma$ continues to decrease until reaching the original point.
The total change in $\gamma$ is $-2 \pi$ and then $\Delta N_c=-1$, corresponding to
$l_+ =0$, $ l_-=1$ in Eq. (\ref{dqb}).

More complex circuits are possible, such as those forming the shape of the number 8, where one critical point is enclosed in the clockwise direction and the other in the counterclockwise direction. This configuration results in a change of
$\Delta N_c= \pm 2$.
Further examples are provided below.

A word of caution is warranted here:
the actual value of the Berry phase as well
as the polarization (charge) at one point depend on definitions and subtleties \cite{Wata18,Aligia23} but Eq. (\ref{dq2}) for the \textit{changes} in the
corresponding quantities is firmly established.

\section{Calculations for some pumping trajectories}

\label{calc}

\begin{figure}[th]
\begin{center}
\includegraphics[width=0.7\columnwidth]{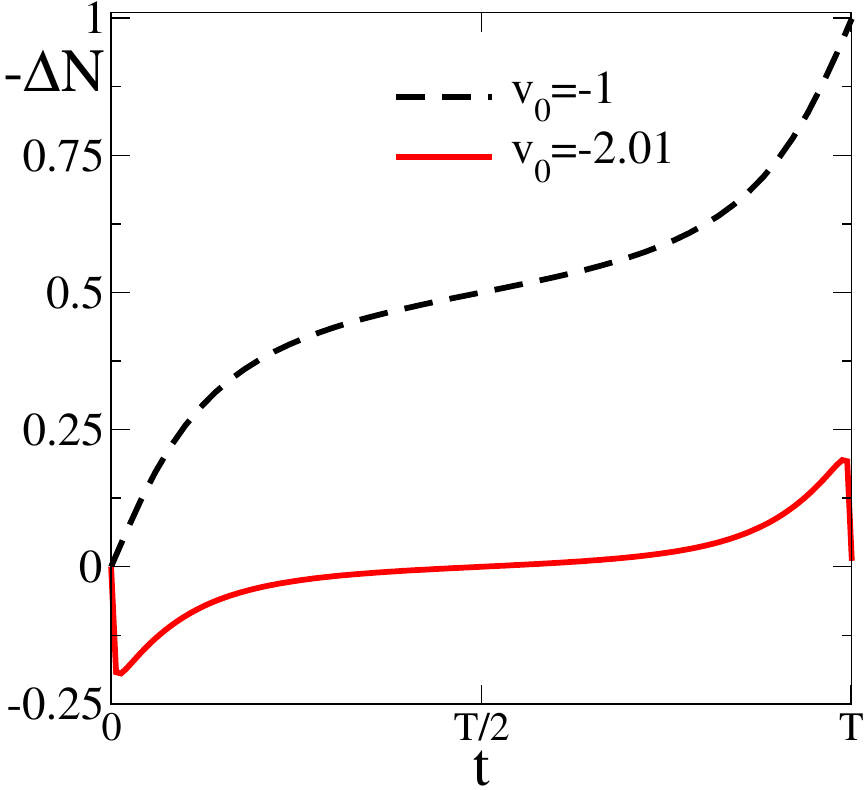}
\includegraphics[width=0.7\columnwidth]{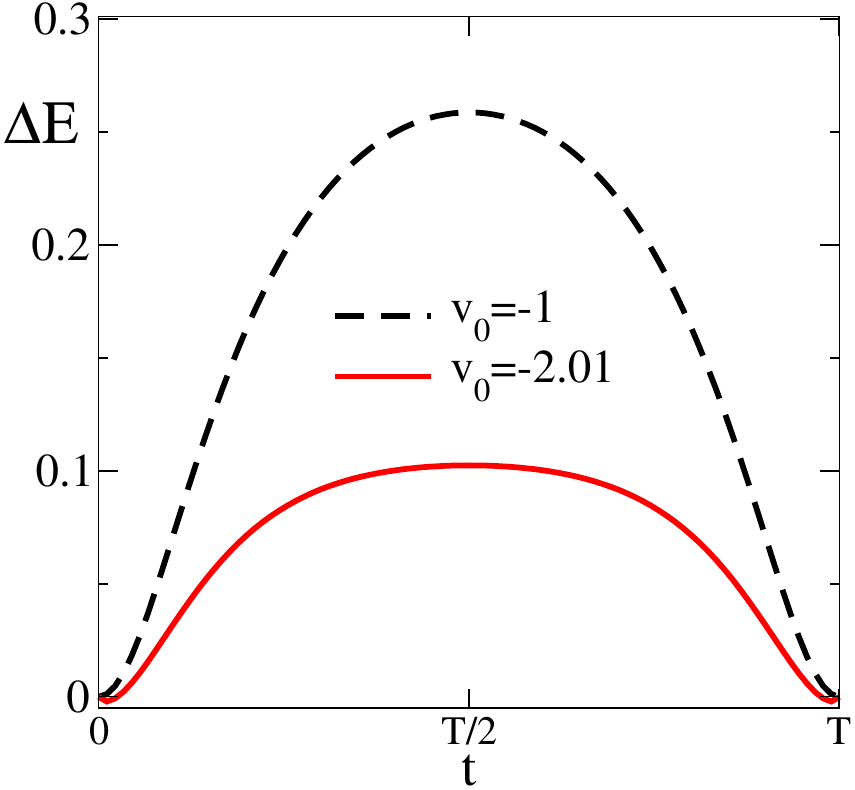}
\end{center}
\caption{Transported number of particles (top) and energy (bottom) as a function of time for two selected values of $v_0$, $\Delta_0=0$,
$v_1=\Delta_1=1$, and zero temperature $\tau=0$.}
\label{DE-vs-t}
\end{figure}

In this section we calculate the particle and heat transport in
the half-filled non-interacting RMC for several closed circuits,
assuming fast thermalization unless otherwise stated.
As it is usual, we choose elliptical pumping trajectories of the form
\begin{eqnarray}\label{pumping-cycle}
 w(t) &=& 1 \nonumber\\
 v(t) &=& v_0+v_1 \mbox{cos}(2 \pi t/T) \nonumber\\
 \Delta(t)&=&\Delta_0 + \Delta_1 \mbox{sin}(2 \pi t/T),
\label{circ}
\end{eqnarray}
where $T$ is the period of the pumping.
We take positive values of $v_1$ and $\Delta_1$ which correspond
to anticlockwise direction in Fig. \ref{berry}.
Note that if we take $\Delta_0=0$, $\Delta(t)$ [$v(t)$] is an odd [even] function of $t$, while its derivative is odd [even]. This
implies that $\Omega_1(t)$ is even [see Eq. (\ref{ome3})]. As a consequence, the contribution to the integral in the expression of the
energy current Eq. (\ref{det}) at time $t$ is exactly the opposite
of that at time $T-t$, and therefore the total transported
energy current for an integer number of cycles vanishes. The same
happens with the heat current in a half-filled system since
$\mu=0$ in this case.

In Fig. \ref{DE-vs-t} we show the evolution of particle and energy transport in two of these elliptical circuits.
To facilitate comparison between both quantities, the
negative of the charge transported to the right is plotted (which 
coincides with the energy gain at the left of the link),
For $v_0=-1$, only the
critical point $(v,\Delta)=(-1,0)$ is contained inside the circuit,
and $l_-=1$, $l_+=0$. Therefore
using Eq. (\ref{dqb}) $\Delta N_c=\Delta N(T)=-1$,
in agreement with
the value $\Delta N(T)=-1$ observed in the figure. For intermediate
times, $\Delta N(t)$ can be inferred from Fig. \ref{berry}. The Berry phase stars at $\pi$ for $t=0$ for which $(v,\Delta)=(0,0)$.
At $t=T/4$ the system is at the point $(v,\Delta)=(-1,1)$ for which $\gamma \sim \pi/2$
[corresponding to $\Delta N=-1/4$ according to Eq. (\ref{dq2})].
At $t=T/2$ the system is at the point (-2,0) for which $\gamma=0$ 
($\equiv 2 \pi$),
implying $\Delta N=-1/2$. For larger times $\gamma$ continues decreasing
reaching the point $\gamma=- \pi \equiv \pi$ at $t=T$.

In a similar way one can analyze the energy transport for $v_0=1$.
The main difference is that $\Delta$ changes sign at $t=T/2$ and
therefore, the derivative changes sign and, as explained above,
the transported energy vanishes at $t=T$.

For $v_0=-2.01$, both critical points $(v,\Delta)=(\pm 1,0)$ are
outside the closed circuit, $l_+=l_-=0$, and then from Eq. (\ref{dqb})
$\Delta N_c=\Delta N(T)=0$, as observed. As before, the contour
plot of the Berry phase (Fig. \ref{berry}) provides an understanding
of the particle and energy transport. Since the starting point
of the cycle (which coincides with the end point) is near the critical
point $(v,\Delta)=(1,0)$, the Berry phase changes more rapidly with time at
$t=0$ mod $T$  and as a consequence, the  corresponding changes in $\Delta N(t)$ are also more pronounced.

\begin{figure}[th]
\begin{center}
\includegraphics[width=0.7\columnwidth]{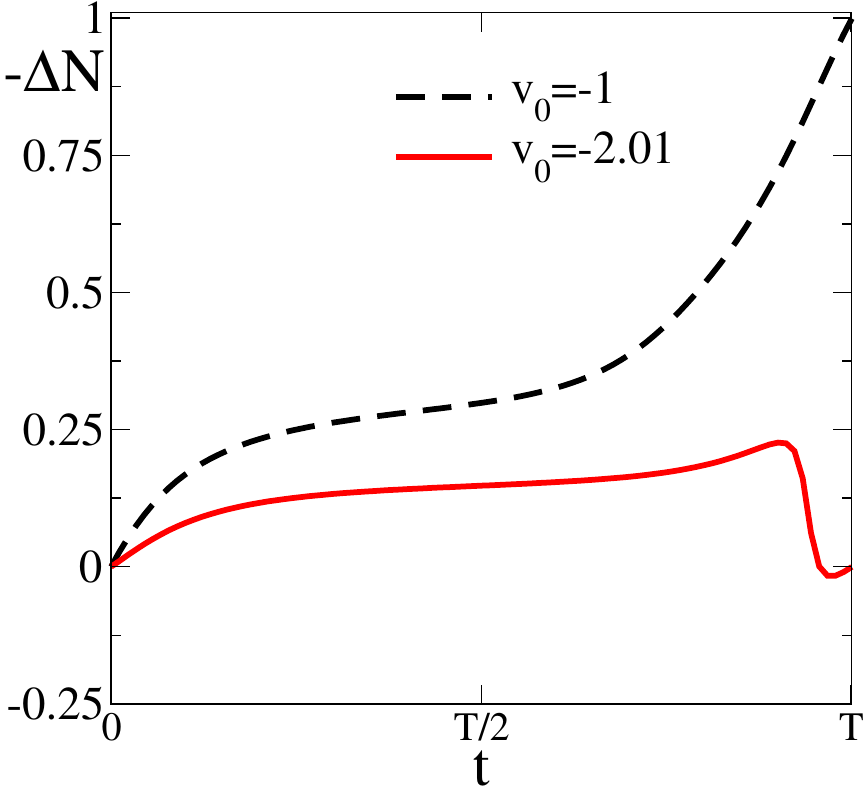}
\includegraphics[width=0.7\columnwidth]{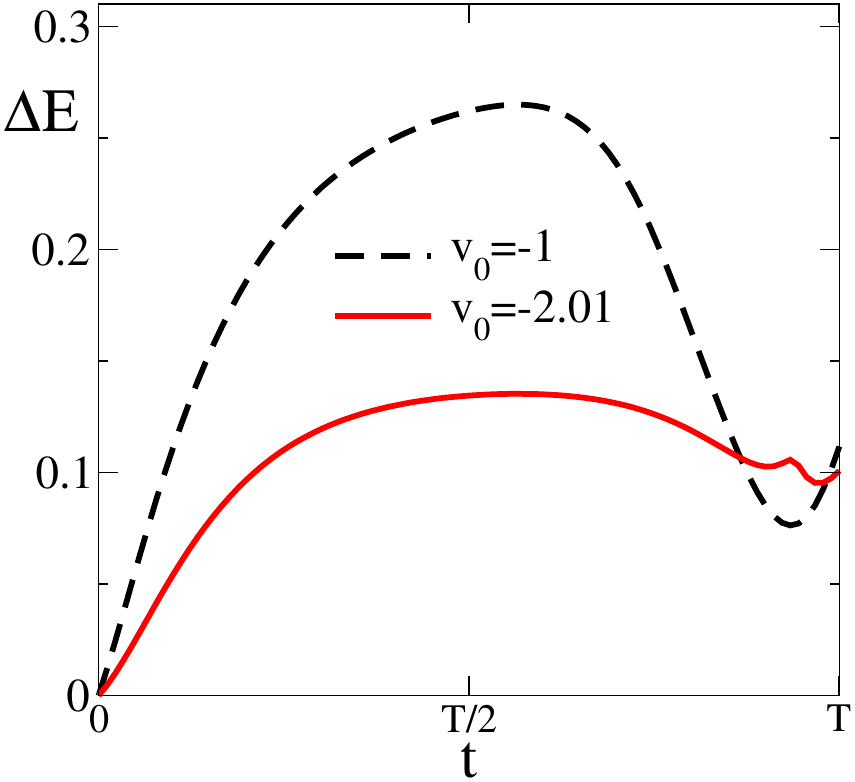}
\end{center}
\caption{Same as Fig. \ref{DE-vs-t} for  $\Delta_0=0.4$.}
\label{DE-vs-t-vs-d0}
\end{figure}

To break the symmetry $t \rightarrow T-t$ and allow for energy pumping
in the cycle we displace the center of the ellipse to $\Delta_0=0.4$.
The corresponding results for the transported charge and heat are
represented in Fig. \ref{DE-vs-t-vs-d0}. Qualitatively the results
are similar as before. However, the steep change in $\Delta N$ with
increasing time is displaced to a time slightly lower than $T$ [where
the trajectory passes near the critical point $(v,\Delta)=(1,0)$],
and as expected, there is a finite amount of energy transfer in the
complete cycle.

\begin{figure}
    \centering
    \includegraphics[width=0.9\linewidth]{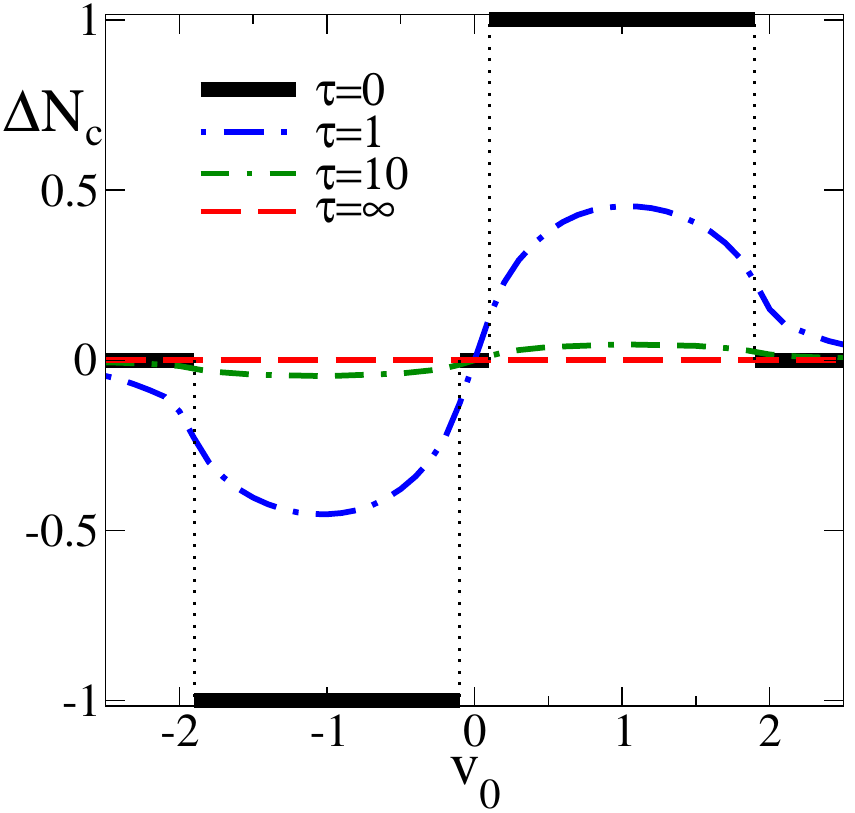}
    \caption{Total transported charge in the period $\Delta N(T)$
    as a function of $v_0$ for several temperatures $\tau$ and
    $\Delta_0=0$, $v_1=0.9$, $\Delta_1=1$.}
    \label{topo}
\end{figure}

From the above results and discussion, it is clear that $\Delta N(T)$
is quantized and has topological character, in contrast to
$\Delta E(T)$. In Fig. \ref{topo} we illustrate how the temperature
affects the topological transitions as a function of $v_0$.
First we discuss the behavior at zero temperature.
For large and negative $v_0$, $l_+ = l_-=0$ and $\Delta N_c=\Delta N(T)=0$
[see Eq. (\ref{dqb}].
Increasing $v_0$, at $v_0=-1.9$ the circuit is displaced to include
the critical point at $(v,\Delta)=(-1,0)$, causing a topological
transition a jump in $l_-$ from 0 to 1 and [from Eq. ((\ref{dqb})] a jump in  $\Delta N_c$ from 0 to -1. At $v_0=-0.1$, the point $(-1,0)$ leaves the circuit causing that $l_-$ and $\Delta N_c$ return again to 0. 
At $v_0=0.1$, the point $(1,0)$ enters the circuit, implying a jump in $l_+$ and $\Delta N_c$ from 0 to 1.
Finally for $v_0>1.9$ both critical points lie again
out of the circuit and $l_+ = l_-=\Delta N_c=0$.

The effect of temperature is to transform the transition
in a smooth crossover. This might be relevant for experiments with cold
atoms in which some heating is unavoidable \cite{Naka16,Lohse16,Walter23,Viebahn23}.

\begin{figure}[th]
\begin{center}
\includegraphics[width=0.7\columnwidth]{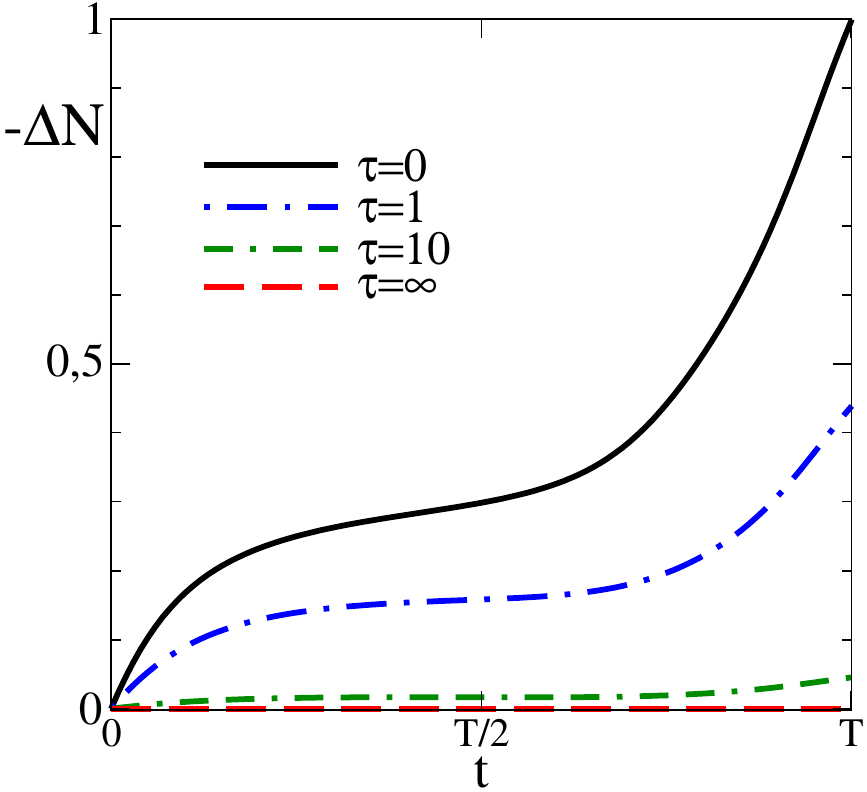}
\includegraphics[width=0.7\columnwidth]{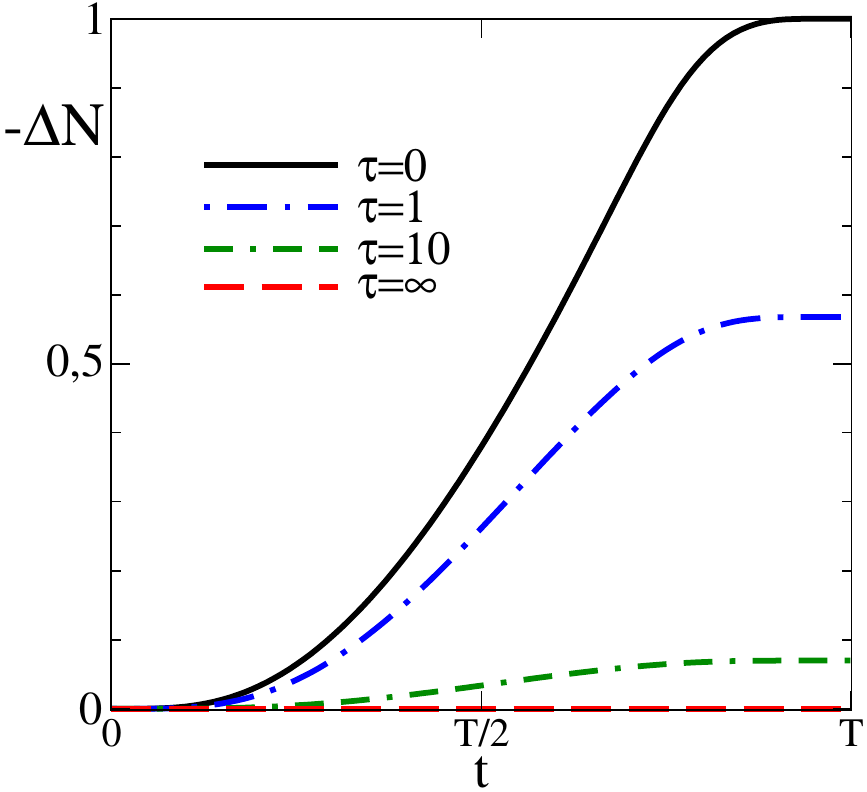}
\end{center}
\caption{Transported charge between two unit
cells (top) or between two atoms in the same unit cell (bottom) as a function of time for $v_0=-1$, $\Delta_0=0.4$,
$v_1=\Delta_1=1$, and several temperatures $\tau$.}
\label{2n}
\end{figure}

In Fig. \ref{2n} we show how the transported charge between two
unit cells evolves with time at different temperatures, and compare
it with the corresponding result between
two sites in the same unit cell (obtained integrating
$\langle \tilde{J}_{p\sigma } \rangle$  instead of
$\langle J_{p\sigma } \rangle$, see the end of Section \ref{model}).
While in both cases the transported number of particles are different at intermediate times, they coincide at zero temperature for $t=T$,
reflecting its topological nature. As expected from the results
of Fig. \ref{topo}, the effect of temperature is to reduce the
transported number of particles inside the topological region.
For very large temperature, the transported charge tends to
zero, as explained at the end of Section \ref{geom}.

\begin{figure}[th]
\begin{center}
\includegraphics[width=0.7\columnwidth]{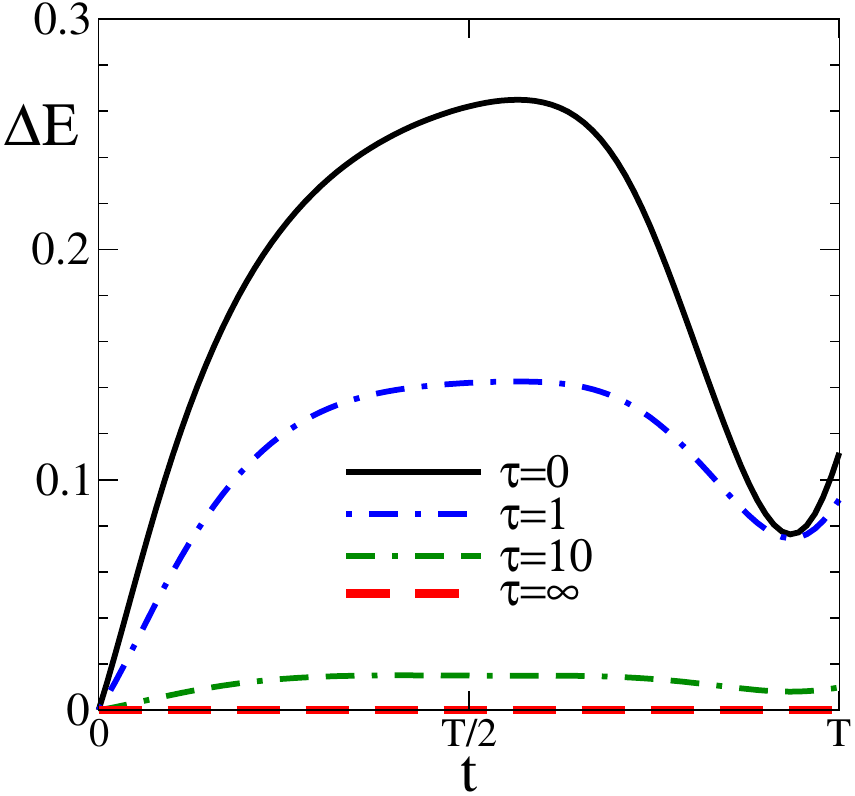}
\includegraphics[width=0.7\columnwidth]{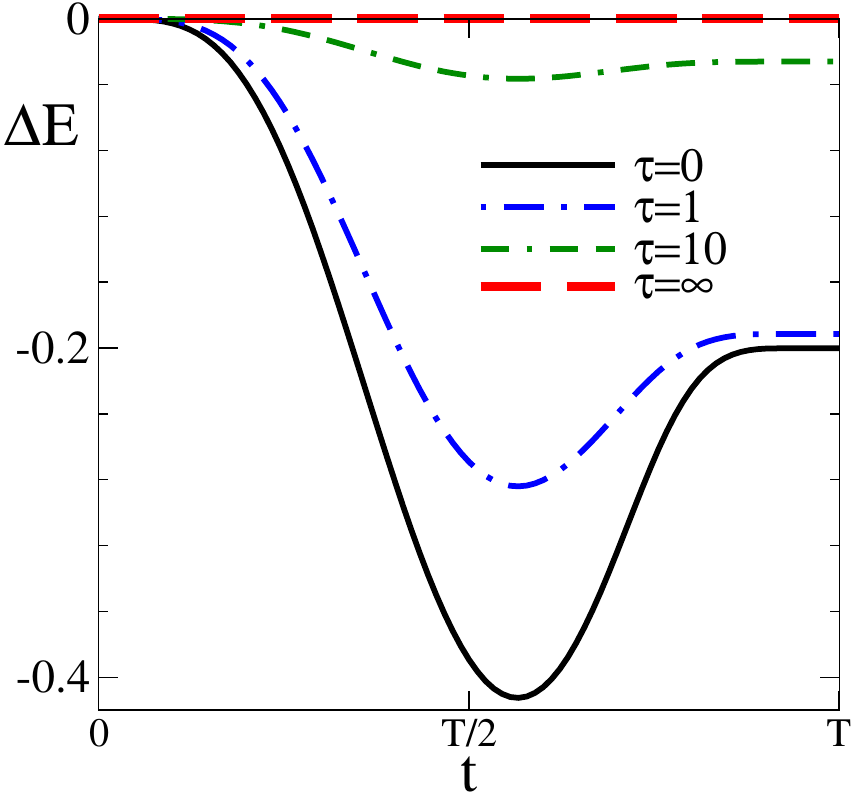}
\end{center}
\caption{Transported energy between two unit
cells (top) or between two atoms in the same unit cell (bottom) as a function of time for $v_0=-1$, $\Delta_0=0.4$,
$v_1=\Delta_1=1$, and several temperatures $\tau$.}
\label{2e}
\end{figure}

In Fig. \ref{2e} we represent the corresponding transported
energy (which coincides with the transported heat since $\mu=0$)
for the same circuit and parameters as above. While the effect
of temperature in reducing the transported energy is expected
and qualitatively similar as for the charge transport, note that
the current between the two links is different, even at the
end of the cycle [$\Delta E(T) \neq \Delta \tilde{E}(T)$]. This
implies that repeating the cycle, there is a net transport
of energy among even and odd sites of the chain
and the environment, except for some usual symmetric cycles
($\Delta_0=0$ in our case, as discussed at the beginning of this Section).

\begin{figure}[bh]
\begin{center}
\includegraphics[width=0.7\columnwidth]
{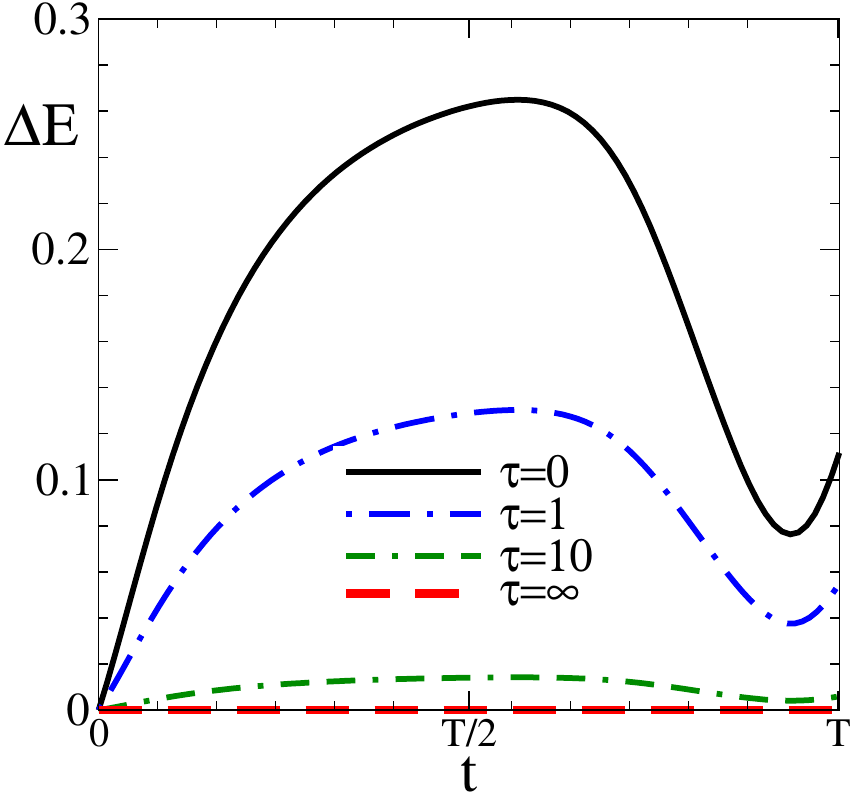}
\end{center}
\caption{Same as Fig. \ref{2e} top for slow thermalization.}
\label{slow}
\end{figure}

In Fig. \ref{slow} we show the corresponding result for the transported 
energy and heat between two unit cells for slow thermalization, to compare
it with the case shown in Fig. \ref{2e} top, which corresponds to fast
thermalization. One can see that the results are very similar, except for small
differences at temperatures $\tau \sim 1$.

\section{Summary and discussion}

\label{sum}

We have studied particle (charge) and energy and heat transport
under an adiabatic change of the parameters of  the RMC focusing
in the non-interacting case $U=0$. While for a cyclic
evolution at zero temperature and half filling, the transported
number of particles in a period $\Delta N(T)$ is known to
be proportional to the change in the Berry phase for any $U$ [Eq. (\ref{dq2})], we provide for $U=0$, general expressions for
the transported charge, energy and heat currents, for any time of
an adiabatic pumping trajectory, filling
and temperature, as double integrals of analytical functions
[Eqs. (\ref{ome3}) and (\ref{fi}) to (\ref{det})]. These expressions
correspond to transport between two neighboring unit cells.
The corresponding expressions for transport between two sites
in the same unit cell are obtained
interchanging $v\leftrightarrow w$ and changing the sign of $\Delta$.

From the geometrical character of the integral we have shown
that for a completely filled system or at infinite temperature,
all transported quantities vanish for any time $t$.
For the completely filled system, this result is valid
also in the interacting case $U \neq 0$. We expect that the
effect of temperature is also generally valid.

We have evaluated explicitly the transported charge and heat
for several simple pumping cycles at half filling and several
temperatures. Notably for some usual symmetric
elliptical pumping cycles used in the literature,
the transported energy vanishes identically, regardless of whether a critical point is enclosed within the trajectory.
For general cycles, energy is transferred between the odd
and even sites of the chain.

Only the charge current at half filling and zero temperature
has a topological character for an integer number of cycles.
In general as the temperature increases, the transported
charge and energy decrease.

Although our results are strictly valid for very long cycle periods, in practice, they apply when the frequency of the pumping cycle is much smaller than the energy gap throughout the entire trajectory.

For finite $U$, the energy transport can be evaluated integrating
the expectation value of the heat current [Eqs. (\ref{jp2}) and (\ref{je})]
using infinite time-evolving block
decimation \cite{Bertok22}.

\section*{Acknowledgments}

We thank Liliana Arrachea for helpfull discussions.
AAA acknowledges financial support provided by PICT 2020A 03661 of the Agencia I+D+i, Argentina.

\bibliography{ref.bib}

\providecommand{\noopsort}[1]{}\providecommand{\singleletter}[1]{#1}%
\begin{thebibliography}{44}%
\makeatletter
\providecommand \@ifxundefined [1]{%
 \@ifx{#1\undefined}
}%
\providecommand \@ifnum [1]{%
 \ifnum #1\expandafter \@firstoftwo
 \else \expandafter \@secondoftwo
 \fi
}%
\providecommand \@ifx [1]{%
 \ifx #1\expandafter \@firstoftwo
 \else \expandafter \@secondoftwo
 \fi
}%
\providecommand \natexlab [1]{#1}%
\providecommand \enquote  [1]{``#1''}%
\providecommand \bibnamefont  [1]{#1}%
\providecommand \bibfnamefont [1]{#1}%
\providecommand \citenamefont [1]{#1}%
\providecommand \href@noop [0]{\@secondoftwo}%
\providecommand \href [0]{\begingroup \@sanitize@url \@href}%
\providecommand \@href[1]{\@@startlink{#1}\@@href}%
\providecommand \@@href[1]{\endgroup#1\@@endlink}%
\providecommand \@sanitize@url [0]{\catcode `\\12\catcode `\$12\catcode
  `\&12\catcode `\#12\catcode `\^12\catcode `\_12\catcode `\%12\relax}%
\providecommand \@@startlink[1]{}%
\providecommand \@@endlink[0]{}%
\providecommand \url  [0]{\begingroup\@sanitize@url \@url }%
\providecommand \@url [1]{\endgroup\@href {#1}{\urlprefix }}%
\providecommand \urlprefix  [0]{URL }%
\providecommand \Eprint [0]{\href }%
\providecommand \doibase [0]{http://dx.doi.org/}%
\providecommand \selectlanguage [0]{\@gobble}%
\providecommand \bibinfo  [0]{\@secondoftwo}%
\providecommand \bibfield  [0]{\@secondoftwo}%
\providecommand \translation [1]{[#1]}%
\providecommand \BibitemOpen [0]{}%
\providecommand \bibitemStop [0]{}%
\providecommand \bibitemNoStop [0]{.\EOS\space}%
\providecommand \EOS [0]{\spacefactor3000\relax}%
\providecommand \BibitemShut  [1]{\csname bibitem#1\endcsname}%
\let\auto@bib@innerbib\@empty
\bibitem [{\citenamefont {Ando}(2013)}]{Ando13}%
  \BibitemOpen
  \bibfield  {author} {\bibinfo {author} {\bibfnamefont {Y.}~\bibnamefont
  {Ando}},\ }\href@noop {} {\bibfield  {journal} {\bibinfo  {journal} {Journal
  of the Physical Society of Japan}\ }\textbf {\bibinfo {volume} {82}},\
  \bibinfo {pages} {102001} (\bibinfo {year} {2013})}\BibitemShut {NoStop}%
\bibitem [{\citenamefont {Bradlyn}\ \emph {et~al.}(2017)\citenamefont
  {Bradlyn}, \citenamefont {Elcoro}, \citenamefont {Cano}, \citenamefont
  {Vergniory}, \citenamefont {Wang}, \citenamefont {Felser}, \citenamefont
  {Aroyo},\ and\ \citenamefont {Bernevig}}]{Bradlyn17}%
  \BibitemOpen
  \bibfield  {author} {\bibinfo {author} {\bibfnamefont {B.}~\bibnamefont
  {Bradlyn}}, \bibinfo {author} {\bibfnamefont {L.}~\bibnamefont {Elcoro}},
  \bibinfo {author} {\bibfnamefont {J.}~\bibnamefont {Cano}}, \bibinfo {author}
  {\bibfnamefont {M.~G.}\ \bibnamefont {Vergniory}}, \bibinfo {author}
  {\bibfnamefont {Z.}~\bibnamefont {Wang}}, \bibinfo {author} {\bibfnamefont
  {C.}~\bibnamefont {Felser}}, \bibinfo {author} {\bibfnamefont {M.~I.}\
  \bibnamefont {Aroyo}}, \ and\ \bibinfo {author} {\bibfnamefont {B.~A.}\
  \bibnamefont {Bernevig}},\ }\href {\doibase 10.1038/nature23268} {\bibfield
  {journal} {\bibinfo  {journal} {Nature}\ }\textbf {\bibinfo {volume} {547}},\
  \bibinfo {pages} {298} (\bibinfo {year} {2017})}\BibitemShut {NoStop}%
\bibitem [{\citenamefont {Sato}\ and\ \citenamefont {Ando}(2017)}]{Sato17}%
  \BibitemOpen
  \bibfield  {author} {\bibinfo {author} {\bibfnamefont {M.}~\bibnamefont
  {Sato}}\ and\ \bibinfo {author} {\bibfnamefont {Y.}~\bibnamefont {Ando}},\
  }\href {\doibase 10.1088/1361-6633/aa6ac7} {\bibfield  {journal} {\bibinfo
  {journal} {Reports on Progress in Physics}\ }\textbf {\bibinfo {volume}
  {80}},\ \bibinfo {pages} {076501} (\bibinfo {year} {2017})}\BibitemShut
  {NoStop}%
\bibitem [{\citenamefont {Berry}(1984)}]{Berry84}%
  \BibitemOpen
  \bibfield  {author} {\bibinfo {author} {\bibfnamefont {M.~V.}\ \bibnamefont
  {Berry}},\ }\href@noop {} {\bibfield  {journal} {\bibinfo  {journal} {Proc.
  R. Soc. Lond. A}\ }\textbf {\bibinfo {volume} {392}},\ \bibinfo {pages} {45}
  (\bibinfo {year} {1984})}\BibitemShut {NoStop}%
\bibitem [{\citenamefont {Thouless}(1983)}]{Thou83}%
  \BibitemOpen
  \bibfield  {author} {\bibinfo {author} {\bibfnamefont {D.~J.}\ \bibnamefont
  {Thouless}},\ }\href {\doibase 10.1103/PhysRevB.27.6083} {\bibfield
  {journal} {\bibinfo  {journal} {Phys. Rev. B}\ }\textbf {\bibinfo {volume}
  {27}},\ \bibinfo {pages} {6083} (\bibinfo {year} {1983})}\BibitemShut
  {NoStop}%
\bibitem [{\citenamefont {Niu}\ and\ \citenamefont {Thouless}(1984)}]{Niu84}%
  \BibitemOpen
  \bibfield  {author} {\bibinfo {author} {\bibfnamefont {Q.}~\bibnamefont
  {Niu}}\ and\ \bibinfo {author} {\bibfnamefont {D.~J.}\ \bibnamefont
  {Thouless}},\ }\href {\doibase 10.1088/0305-4470/17/12/016} {\bibfield
  {journal} {\bibinfo  {journal} {J. Phys. A: Math. Gen.}\ }\textbf {\bibinfo
  {volume} {17}},\ \bibinfo {pages} {2453} (\bibinfo {year}
  {1984})}\BibitemShut {NoStop}%
\bibitem [{\citenamefont {Wang}\ \emph {et~al.}(2013)\citenamefont {Wang},
  \citenamefont {Troyer},\ and\ \citenamefont {Dai}}]{Wang13}%
  \BibitemOpen
  \bibfield  {author} {\bibinfo {author} {\bibfnamefont {L.}~\bibnamefont
  {Wang}}, \bibinfo {author} {\bibfnamefont {M.}~\bibnamefont {Troyer}}, \ and\
  \bibinfo {author} {\bibfnamefont {X.}~\bibnamefont {Dai}},\ }\href {\doibase
  10.1103/PhysRevLett.111.026802} {\bibfield  {journal} {\bibinfo  {journal}
  {Phys. Rev. Lett.}\ }\textbf {\bibinfo {volume} {111}},\ \bibinfo {pages}
  {026802} (\bibinfo {year} {2013})}\BibitemShut {NoStop}%
\bibitem [{\citenamefont {Citro}\ and\ \citenamefont
  {Aidelsburger}(2023)}]{Citro23}%
  \BibitemOpen
  \bibfield  {author} {\bibinfo {author} {\bibfnamefont {R.}~\bibnamefont
  {Citro}}\ and\ \bibinfo {author} {\bibfnamefont {M.}~\bibnamefont
  {Aidelsburger}},\ }\href {\doibase 10.1038/s42254-022-00545-0} {\bibfield
  {journal} {\bibinfo  {journal} {Nature Reviews Physics}\ }\textbf {\bibinfo
  {volume} {5}},\ \bibinfo {pages} {87} (\bibinfo {year} {2023})}\BibitemShut
  {NoStop}%
\bibitem [{\citenamefont {Zak}(1989)}]{Zak89}%
  \BibitemOpen
  \bibfield  {author} {\bibinfo {author} {\bibfnamefont {J.}~\bibnamefont
  {Zak}},\ }\href {\doibase 10.1103/PhysRevLett.62.2747} {\bibfield  {journal}
  {\bibinfo  {journal} {Phys. Rev. Lett.}\ }\textbf {\bibinfo {volume} {62}},\
  \bibinfo {pages} {2747} (\bibinfo {year} {1989})}\BibitemShut {NoStop}%
\bibitem [{\citenamefont {Tewari}\ and\ \citenamefont {Sau}(2012)}]{Tewari12}%
  \BibitemOpen
  \bibfield  {author} {\bibinfo {author} {\bibfnamefont {S.}~\bibnamefont
  {Tewari}}\ and\ \bibinfo {author} {\bibfnamefont {J.~D.}\ \bibnamefont
  {Sau}},\ }\href {\doibase 10.1103/PhysRevLett.109.150408} {\bibfield
  {journal} {\bibinfo  {journal} {Phys. Rev. Lett.}\ }\textbf {\bibinfo
  {volume} {109}},\ \bibinfo {pages} {150408} (\bibinfo {year}
  {2012})}\BibitemShut {NoStop}%
\bibitem [{\citenamefont {Asbóth}\ \emph {et~al.}(2016)\citenamefont
  {Asbóth}, \citenamefont {Oroszlány},\ and\ \citenamefont {Pályi}}]{Asb16}%
  \BibitemOpen
  \bibfield  {author} {\bibinfo {author} {\bibfnamefont {J.~K.}\ \bibnamefont
  {Asbóth}}, \bibinfo {author} {\bibfnamefont {L.}~\bibnamefont {Oroszlány}},
  \ and\ \bibinfo {author} {\bibfnamefont {A.}~\bibnamefont {Pályi}},\ }\href
  {\doibase 10.1007/978-3-319-25607-8} {\emph {\bibinfo {title} {A Short Course
  on Topological Insulators}}}\ (\bibinfo  {publisher} {Springer International
  Publishing},\ \bibinfo {year} {2016})\BibitemShut {NoStop}%
\bibitem [{\citenamefont {Cardano}\ \emph {et~al.}(2017)\citenamefont
  {Cardano}, \citenamefont {D'Errico}, \citenamefont {Dauphin}, \citenamefont
  {Maffei}, \citenamefont {Piccirillo}, \citenamefont {de~Lisio}, \citenamefont
  {De~Filippis}, \citenamefont {Cataudella}, \citenamefont {Santamato},
  \citenamefont {Marrucci}, \citenamefont {Lewenstein},\ and\ \citenamefont
  {Massignan}}]{Cardano17}%
  \BibitemOpen
  \bibfield  {author} {\bibinfo {author} {\bibfnamefont {F.}~\bibnamefont
  {Cardano}}, \bibinfo {author} {\bibfnamefont {A.}~\bibnamefont {D'Errico}},
  \bibinfo {author} {\bibfnamefont {A.}~\bibnamefont {Dauphin}}, \bibinfo
  {author} {\bibfnamefont {M.}~\bibnamefont {Maffei}}, \bibinfo {author}
  {\bibfnamefont {B.}~\bibnamefont {Piccirillo}}, \bibinfo {author}
  {\bibfnamefont {C.}~\bibnamefont {de~Lisio}}, \bibinfo {author}
  {\bibfnamefont {G.}~\bibnamefont {De~Filippis}}, \bibinfo {author}
  {\bibfnamefont {V.}~\bibnamefont {Cataudella}}, \bibinfo {author}
  {\bibfnamefont {E.}~\bibnamefont {Santamato}}, \bibinfo {author}
  {\bibfnamefont {L.}~\bibnamefont {Marrucci}}, \bibinfo {author}
  {\bibfnamefont {M.}~\bibnamefont {Lewenstein}}, \ and\ \bibinfo {author}
  {\bibfnamefont {P.}~\bibnamefont {Massignan}},\ }\href {\doibase
  10.1038/ncomms15516} {\bibfield  {journal} {\bibinfo  {journal} {Nature
  Communications}\ }\textbf {\bibinfo {volume} {8}},\ \bibinfo {pages} {15516}
  (\bibinfo {year} {2017})}\BibitemShut {NoStop}%
\bibitem [{\citenamefont {P\'erez~Daroca}\ and\ \citenamefont
  {Aligia}(2021)}]{Daroca21}%
  \BibitemOpen
  \bibfield  {author} {\bibinfo {author} {\bibfnamefont {D.}~\bibnamefont
  {P\'erez~Daroca}}\ and\ \bibinfo {author} {\bibfnamefont {A.~A.}\
  \bibnamefont {Aligia}},\ }\href {\doibase 10.1103/PhysRevB.104.115125}
  {\bibfield  {journal} {\bibinfo  {journal} {Phys. Rev. B}\ }\textbf {\bibinfo
  {volume} {104}},\ \bibinfo {pages} {115125} (\bibinfo {year}
  {2021})}\BibitemShut {NoStop}%
\bibitem [{\citenamefont {Ostahie}\ \emph {et~al.}(2023)\citenamefont
  {Ostahie}, \citenamefont {Sticlet}, \citenamefont {Moca}, \citenamefont
  {D\'ora}, \citenamefont {Werner}, \citenamefont {Asb\'oth},\ and\
  \citenamefont {Zar\'and}}]{Osta23}%
  \BibitemOpen
  \bibfield  {author} {\bibinfo {author} {\bibfnamefont {B.}~\bibnamefont
  {Ostahie}}, \bibinfo {author} {\bibfnamefont {D.}~\bibnamefont {Sticlet}},
  \bibinfo {author} {\bibfnamefont {C.~P.}\ \bibnamefont {Moca}}, \bibinfo
  {author} {\bibfnamefont {B.}~\bibnamefont {D\'ora}}, \bibinfo {author}
  {\bibfnamefont {M.~A.}\ \bibnamefont {Werner}}, \bibinfo {author}
  {\bibfnamefont {J.~K.}\ \bibnamefont {Asb\'oth}}, \ and\ \bibinfo {author}
  {\bibfnamefont {G.}~\bibnamefont {Zar\'and}},\ }\href {\doibase
  10.1103/PhysRevB.108.035126} {\bibfield  {journal} {\bibinfo  {journal}
  {Phys. Rev. B}\ }\textbf {\bibinfo {volume} {108}},\ \bibinfo {pages}
  {035126} (\bibinfo {year} {2023})}\BibitemShut {NoStop}%
\bibitem [{\citenamefont {Aligia}(2023)}]{Aligia23}%
  \BibitemOpen
  \bibfield  {author} {\bibinfo {author} {\bibfnamefont {A.~A.}\ \bibnamefont
  {Aligia}},\ }\href {\doibase 10.1103/PhysRevB.107.075153} {\bibfield
  {journal} {\bibinfo  {journal} {Phys. Rev. B}\ }\textbf {\bibinfo {volume}
  {107}},\ \bibinfo {pages} {075153} (\bibinfo {year} {2023})}\BibitemShut
  {NoStop}%
\bibitem [{\citenamefont {Resta}(1994)}]{Resta94}%
  \BibitemOpen
  \bibfield  {author} {\bibinfo {author} {\bibfnamefont {R.}~\bibnamefont
  {Resta}},\ }\href {\doibase 10.1103/RevModPhys.66.899} {\bibfield  {journal}
  {\bibinfo  {journal} {Rev. Mod. Phys.}\ }\textbf {\bibinfo {volume} {66}},\
  \bibinfo {pages} {899} (\bibinfo {year} {1994})}\BibitemShut {NoStop}%
\bibitem [{\citenamefont {Xiao}\ \emph {et~al.}(2010)\citenamefont {Xiao},
  \citenamefont {Chang},\ and\ \citenamefont {Niu}}]{Xiao10}%
  \BibitemOpen
  \bibfield  {author} {\bibinfo {author} {\bibfnamefont {D.}~\bibnamefont
  {Xiao}}, \bibinfo {author} {\bibfnamefont {M.-C.}\ \bibnamefont {Chang}}, \
  and\ \bibinfo {author} {\bibfnamefont {Q.}~\bibnamefont {Niu}},\ }\href
  {\doibase 10.1103/RevModPhys.82.1959} {\bibfield  {journal} {\bibinfo
  {journal} {Rev. Mod. Phys.}\ }\textbf {\bibinfo {volume} {82}},\ \bibinfo
  {pages} {1959} (\bibinfo {year} {2010})}\BibitemShut {NoStop}%
\bibitem [{\citenamefont {Vanderbilt}(2018)}]{Vander18}%
  \BibitemOpen
  \bibfield  {author} {\bibinfo {author} {\bibfnamefont {D.}~\bibnamefont
  {Vanderbilt}},\ }\href@noop {} {\emph {\bibinfo {title} {Berry Phases in
  Electronic Structure Theory: Electric Polarization, Orbital Magnetization and
  Topological Insulators}}}\ (\bibinfo  {publisher} {Cambridge University
  Press},\ \bibinfo {year} {2018})\BibitemShut {NoStop}%
\bibitem [{\citenamefont {Bradlyn}\ and\ \citenamefont
  {Iraola}(2022)}]{Bradlyn22}%
  \BibitemOpen
  \bibfield  {author} {\bibinfo {author} {\bibfnamefont {B.}~\bibnamefont
  {Bradlyn}}\ and\ \bibinfo {author} {\bibfnamefont {M.}~\bibnamefont
  {Iraola}},\ }\href {\doibase 10.21468/SciPostPhysLectNotes.51} {\bibfield
  {journal} {\bibinfo  {journal} {SciPost Phys. Lect. Notes}\ ,\ \bibinfo
  {pages} {51}} (\bibinfo {year} {2022})}\BibitemShut {NoStop}%
\bibitem [{\citenamefont {Ortiz}\ and\ \citenamefont {Martin}(1994)}]{Ortiz94}%
  \BibitemOpen
  \bibfield  {author} {\bibinfo {author} {\bibfnamefont {G.}~\bibnamefont
  {Ortiz}}\ and\ \bibinfo {author} {\bibfnamefont {R.~M.}\ \bibnamefont
  {Martin}},\ }\href {\doibase 10.1103/PhysRevB.49.14202} {\bibfield  {journal}
  {\bibinfo  {journal} {Phys. Rev. B}\ }\textbf {\bibinfo {volume} {49}},\
  \bibinfo {pages} {14202} (\bibinfo {year} {1994})}\BibitemShut {NoStop}%
\bibitem [{\citenamefont {Resta}\ and\ \citenamefont {Sorella}(1995)}]{resor}%
  \BibitemOpen
  \bibfield  {author} {\bibinfo {author} {\bibfnamefont {R.}~\bibnamefont
  {Resta}}\ and\ \bibinfo {author} {\bibfnamefont {S.}~\bibnamefont
  {Sorella}},\ }\href {\doibase 10.1103/PhysRevLett.74.4738} {\bibfield
  {journal} {\bibinfo  {journal} {Phys. Rev. Lett.}\ }\textbf {\bibinfo
  {volume} {74}},\ \bibinfo {pages} {4738} (\bibinfo {year}
  {1995})}\BibitemShut {NoStop}%
\bibitem [{\citenamefont {Ortiz}\ \emph {et~al.}(1996)\citenamefont {Ortiz},
  \citenamefont {Ordej\'on}, \citenamefont {Martin},\ and\ \citenamefont
  {Chiappe}}]{oc}%
  \BibitemOpen
  \bibfield  {author} {\bibinfo {author} {\bibfnamefont {G.}~\bibnamefont
  {Ortiz}}, \bibinfo {author} {\bibfnamefont {P.}~\bibnamefont {Ordej\'on}},
  \bibinfo {author} {\bibfnamefont {R.~M.}\ \bibnamefont {Martin}}, \ and\
  \bibinfo {author} {\bibfnamefont {G.}~\bibnamefont {Chiappe}},\ }\href
  {\doibase 10.1103/PhysRevB.54.13515} {\bibfield  {journal} {\bibinfo
  {journal} {Phys. Rev. B}\ }\textbf {\bibinfo {volume} {54}},\ \bibinfo
  {pages} {13515} (\bibinfo {year} {1996})}\BibitemShut {NoStop}%
\bibitem [{\citenamefont {Song}\ \emph {et~al.}(2021)\citenamefont {Song},
  \citenamefont {He}, \citenamefont {Vishwanath},\ and\ \citenamefont
  {Wang}}]{Song21}%
  \BibitemOpen
  \bibfield  {author} {\bibinfo {author} {\bibfnamefont {X.-Y.}\ \bibnamefont
  {Song}}, \bibinfo {author} {\bibfnamefont {Y.-C.}\ \bibnamefont {He}},
  \bibinfo {author} {\bibfnamefont {A.}~\bibnamefont {Vishwanath}}, \ and\
  \bibinfo {author} {\bibfnamefont {C.}~\bibnamefont {Wang}},\ }\href {\doibase
  10.1103/PhysRevResearch.3.023011} {\bibfield  {journal} {\bibinfo  {journal}
  {Phys. Rev. Res.}\ }\textbf {\bibinfo {volume} {3}},\ \bibinfo {pages}
  {023011} (\bibinfo {year} {2021})}\BibitemShut {NoStop}%
\bibitem [{\citenamefont {Watanabe}\ and\ \citenamefont
  {Oshikawa}(2018)}]{Wata18}%
  \BibitemOpen
  \bibfield  {author} {\bibinfo {author} {\bibfnamefont {H.}~\bibnamefont
  {Watanabe}}\ and\ \bibinfo {author} {\bibfnamefont {M.}~\bibnamefont
  {Oshikawa}},\ }\href {\doibase 10.1103/PhysRevX.8.021065} {\bibfield
  {journal} {\bibinfo  {journal} {Phys. Rev. X}\ }\textbf {\bibinfo {volume}
  {8}},\ \bibinfo {pages} {021065} (\bibinfo {year} {2018})}\BibitemShut
  {NoStop}%
\bibitem [{\citenamefont {Nakajima}\ \emph {et~al.}(2016)\citenamefont
  {Nakajima}, \citenamefont {Tomita}, \citenamefont {Taie}, \citenamefont
  {Ichinose}, \citenamefont {Ozawa}, \citenamefont {Wang}, \citenamefont
  {Troyer},\ and\ \citenamefont {Takahashi}}]{Naka16}%
  \BibitemOpen
  \bibfield  {author} {\bibinfo {author} {\bibfnamefont {S.}~\bibnamefont
  {Nakajima}}, \bibinfo {author} {\bibfnamefont {T.}~\bibnamefont {Tomita}},
  \bibinfo {author} {\bibfnamefont {S.}~\bibnamefont {Taie}}, \bibinfo {author}
  {\bibfnamefont {T.}~\bibnamefont {Ichinose}}, \bibinfo {author}
  {\bibfnamefont {H.}~\bibnamefont {Ozawa}}, \bibinfo {author} {\bibfnamefont
  {L.}~\bibnamefont {Wang}}, \bibinfo {author} {\bibfnamefont {M.}~\bibnamefont
  {Troyer}}, \ and\ \bibinfo {author} {\bibfnamefont {Y.}~\bibnamefont
  {Takahashi}},\ }\href {\doibase 10.1038/nphys3622} {\bibfield  {journal}
  {\bibinfo  {journal} {Nature Physics}\ }\textbf {\bibinfo {volume} {12}},\
  \bibinfo {pages} {296} (\bibinfo {year} {2016})}\BibitemShut {NoStop}%
\bibitem [{\citenamefont {Lohse}\ \emph {et~al.}(2016)\citenamefont {Lohse},
  \citenamefont {Schweizer}, \citenamefont {Zilberberg}, \citenamefont
  {Aidelsburger},\ and\ \citenamefont {Bloch}}]{Lohse16}%
  \BibitemOpen
  \bibfield  {author} {\bibinfo {author} {\bibfnamefont {M.}~\bibnamefont
  {Lohse}}, \bibinfo {author} {\bibfnamefont {C.}~\bibnamefont {Schweizer}},
  \bibinfo {author} {\bibfnamefont {O.}~\bibnamefont {Zilberberg}}, \bibinfo
  {author} {\bibfnamefont {M.}~\bibnamefont {Aidelsburger}}, \ and\ \bibinfo
  {author} {\bibfnamefont {I.}~\bibnamefont {Bloch}},\ }\href {\doibase
  doi.org/10.1038/nphys3584} {\bibfield  {journal} {\bibinfo  {journal} {Nature
  Physics}\ }\textbf {\bibinfo {volume} {12}},\ \bibinfo {pages} {350}
  (\bibinfo {year} {2016})}\BibitemShut {NoStop}%
\bibitem [{\citenamefont {Walter}\ \emph {et~al.}(2023)\citenamefont {Walter},
  \citenamefont {Zhu}, \citenamefont {G{\"a}chter}, \citenamefont {Minguzzi},
  \citenamefont {Roschinski}, \citenamefont {Sandholzer}, \citenamefont
  {Viebahn},\ and\ \citenamefont {Esslinger}}]{Walter23}%
  \BibitemOpen
  \bibfield  {author} {\bibinfo {author} {\bibfnamefont {A.-S.}\ \bibnamefont
  {Walter}}, \bibinfo {author} {\bibfnamefont {Z.}~\bibnamefont {Zhu}},
  \bibinfo {author} {\bibfnamefont {M.}~\bibnamefont {G{\"a}chter}}, \bibinfo
  {author} {\bibfnamefont {J.}~\bibnamefont {Minguzzi}}, \bibinfo {author}
  {\bibfnamefont {S.}~\bibnamefont {Roschinski}}, \bibinfo {author}
  {\bibfnamefont {K.}~\bibnamefont {Sandholzer}}, \bibinfo {author}
  {\bibfnamefont {K.}~\bibnamefont {Viebahn}}, \ and\ \bibinfo {author}
  {\bibfnamefont {T.}~\bibnamefont {Esslinger}},\ }\href@noop {} {\bibfield
  {journal} {\bibinfo  {journal} {Nature Physics}\ }\textbf {\bibinfo {volume}
  {19}},\ \bibinfo {pages} {1471} (\bibinfo {year} {2023})}\BibitemShut
  {NoStop}%
\bibitem [{\citenamefont {Viebahn}\ \emph {et~al.}(2024)\citenamefont
  {Viebahn}, \citenamefont {Walter}, \citenamefont {Bertok}, \citenamefont
  {Zhu}, \citenamefont {G\"achter}, \citenamefont {Aligia}, \citenamefont
  {Heidrich-Meisner},\ and\ \citenamefont {Esslinger}}]{Viebahn23}%
  \BibitemOpen
  \bibfield  {author} {\bibinfo {author} {\bibfnamefont {K.}~\bibnamefont
  {Viebahn}}, \bibinfo {author} {\bibfnamefont {A.-S.}\ \bibnamefont {Walter}},
  \bibinfo {author} {\bibfnamefont {E.}~\bibnamefont {Bertok}}, \bibinfo
  {author} {\bibfnamefont {Z.}~\bibnamefont {Zhu}}, \bibinfo {author}
  {\bibfnamefont {M.}~\bibnamefont {G\"achter}}, \bibinfo {author}
  {\bibfnamefont {A.~A.}\ \bibnamefont {Aligia}}, \bibinfo {author}
  {\bibfnamefont {F.}~\bibnamefont {Heidrich-Meisner}}, \ and\ \bibinfo
  {author} {\bibfnamefont {T.}~\bibnamefont {Esslinger}},\ }\href {\doibase
  10.1103/PhysRevX.14.021049} {\bibfield  {journal} {\bibinfo  {journal} {Phys.
  Rev. X}\ }\textbf {\bibinfo {volume} {14}},\ \bibinfo {pages} {021049}
  (\bibinfo {year} {2024})}\BibitemShut {NoStop}%
\bibitem [{\citenamefont {Hayward}\ \emph {et~al.}(2018)\citenamefont
  {Hayward}, \citenamefont {Schweizer}, \citenamefont {Lohse}, \citenamefont
  {Aidelsburger},\ and\ \citenamefont {Heidrich-Meisner}}]{Hay18}%
  \BibitemOpen
  \bibfield  {author} {\bibinfo {author} {\bibfnamefont {A.}~\bibnamefont
  {Hayward}}, \bibinfo {author} {\bibfnamefont {C.}~\bibnamefont {Schweizer}},
  \bibinfo {author} {\bibfnamefont {M.}~\bibnamefont {Lohse}}, \bibinfo
  {author} {\bibfnamefont {M.}~\bibnamefont {Aidelsburger}}, \ and\ \bibinfo
  {author} {\bibfnamefont {F.}~\bibnamefont {Heidrich-Meisner}},\ }\href
  {\doibase 10.1103/PhysRevB.98.245148} {\bibfield  {journal} {\bibinfo
  {journal} {Phys. Rev. B}\ }\textbf {\bibinfo {volume} {98}},\ \bibinfo
  {pages} {245148} (\bibinfo {year} {2018})}\BibitemShut {NoStop}%
\bibitem [{\citenamefont {Nakagawa}\ \emph {et~al.}(2018)\citenamefont
  {Nakagawa}, \citenamefont {Yoshida}, \citenamefont {Peters},\ and\
  \citenamefont {Kawakami}}]{Nakag18}%
  \BibitemOpen
  \bibfield  {author} {\bibinfo {author} {\bibfnamefont {M.}~\bibnamefont
  {Nakagawa}}, \bibinfo {author} {\bibfnamefont {T.}~\bibnamefont {Yoshida}},
  \bibinfo {author} {\bibfnamefont {R.}~\bibnamefont {Peters}}, \ and\ \bibinfo
  {author} {\bibfnamefont {N.}~\bibnamefont {Kawakami}},\ }\href {\doibase
  10.1103/PhysRevB.98.115147} {\bibfield  {journal} {\bibinfo  {journal} {Phys.
  Rev. B}\ }\textbf {\bibinfo {volume} {98}},\ \bibinfo {pages} {115147}
  (\bibinfo {year} {2018})}\BibitemShut {NoStop}%
\bibitem [{\citenamefont {Bertok}\ \emph {et~al.}(2022)\citenamefont {Bertok},
  \citenamefont {Heidrich-Meisner},\ and\ \citenamefont {Aligia}}]{Bertok22}%
  \BibitemOpen
  \bibfield  {author} {\bibinfo {author} {\bibfnamefont {E.}~\bibnamefont
  {Bertok}}, \bibinfo {author} {\bibfnamefont {F.}~\bibnamefont
  {Heidrich-Meisner}}, \ and\ \bibinfo {author} {\bibfnamefont {A.~A.}\
  \bibnamefont {Aligia}},\ }\href {\doibase 10.1103/PhysRevB.106.045141}
  {\bibfield  {journal} {\bibinfo  {journal} {Phys. Rev. B}\ }\textbf {\bibinfo
  {volume} {106}},\ \bibinfo {pages} {045141} (\bibinfo {year}
  {2022})}\BibitemShut {NoStop}%
\bibitem [{\citenamefont {Roura-Bas}\ and\ \citenamefont
  {Aligia}(2023)}]{Roura23}%
  \BibitemOpen
  \bibfield  {author} {\bibinfo {author} {\bibfnamefont {P.}~\bibnamefont
  {Roura-Bas}}\ and\ \bibinfo {author} {\bibfnamefont {A.~A.}\ \bibnamefont
  {Aligia}},\ }\href {\doibase 10.1103/PhysRevB.108.115132} {\bibfield
  {journal} {\bibinfo  {journal} {Phys. Rev. B}\ }\textbf {\bibinfo {volume}
  {108}},\ \bibinfo {pages} {115132} (\bibinfo {year} {2023})}\BibitemShut
  {NoStop}%
\bibitem [{\citenamefont {Moreno~Segura}\ \emph {et~al.}(2023)\citenamefont
  {Moreno~Segura}, \citenamefont {Hallberg},\ and\ \citenamefont
  {Aligia}}]{Moreno23}%
  \BibitemOpen
  \bibfield  {author} {\bibinfo {author} {\bibfnamefont {O.~A.}\ \bibnamefont
  {Moreno~Segura}}, \bibinfo {author} {\bibfnamefont {K.}~\bibnamefont
  {Hallberg}}, \ and\ \bibinfo {author} {\bibfnamefont {A.~A.}\ \bibnamefont
  {Aligia}},\ }\href {\doibase 10.1103/PhysRevB.108.195135} {\bibfield
  {journal} {\bibinfo  {journal} {Phys. Rev. B}\ }\textbf {\bibinfo {volume}
  {108}},\ \bibinfo {pages} {195135} (\bibinfo {year} {2023})}\BibitemShut
  {NoStop}%
\bibitem [{\citenamefont {Hattori}\ \emph {et~al.}(2023)\citenamefont
  {Hattori}, \citenamefont {Ishikawa},\ and\ \citenamefont
  {Kaneko}}]{Hattori23}%
  \BibitemOpen
  \bibfield  {author} {\bibinfo {author} {\bibfnamefont {K.}~\bibnamefont
  {Hattori}}, \bibinfo {author} {\bibfnamefont {K.}~\bibnamefont {Ishikawa}}, \
  and\ \bibinfo {author} {\bibfnamefont {Y.}~\bibnamefont {Kaneko}},\ }\href
  {\doibase 10.1103/PhysRevB.107.115401} {\bibfield  {journal} {\bibinfo
  {journal} {Phys. Rev. B}\ }\textbf {\bibinfo {volume} {107}},\ \bibinfo
  {pages} {115401} (\bibinfo {year} {2023})}\BibitemShut {NoStop}%
\bibitem [{\citenamefont {Argüello-Luengo}\ \emph {et~al.}(2024)\citenamefont
  {Argüello-Luengo}, \citenamefont {Mark}, \citenamefont {Ferlaino},
  \citenamefont {Lewenstein}, \citenamefont {Barbiero},\ and\ \citenamefont
  {Julià-Farré}}]{Arg24}%
  \BibitemOpen
  \bibfield  {author} {\bibinfo {author} {\bibfnamefont {J.}~\bibnamefont
  {Argüello-Luengo}}, \bibinfo {author} {\bibfnamefont {M.~J.}\ \bibnamefont
  {Mark}}, \bibinfo {author} {\bibfnamefont {F.}~\bibnamefont {Ferlaino}},
  \bibinfo {author} {\bibfnamefont {M.}~\bibnamefont {Lewenstein}}, \bibinfo
  {author} {\bibfnamefont {L.}~\bibnamefont {Barbiero}}, \ and\ \bibinfo
  {author} {\bibfnamefont {S.}~\bibnamefont {Julià-Farré}},\ }\href {\doibase
  10.22331/q-2024-03-14-1285} {\bibfield  {journal} {\bibinfo  {journal}
  {Quantum}\ }\textbf {\bibinfo {volume} {8}},\ \bibinfo {pages} {1285}
  (\bibinfo {year} {2024})}\BibitemShut {NoStop}%
\bibitem [{\citenamefont {Tada}(2024)}]{Tada24}%
  \BibitemOpen
  \bibfield  {author} {\bibinfo {author} {\bibfnamefont {Y.}~\bibnamefont
  {Tada}},\ }\href@noop {} {\enquote {\bibinfo {title} {Quantized polarization
  in a generalized {Rice}-{Mele} model at arbitrary filling},}\ } (\bibinfo
  {year} {2024}),\ \Eprint {http://arxiv.org/abs/2404.09262} {arXiv:2404.09262
  [cond-mat.str-el]} \BibitemShut {NoStop}%
\bibitem [{\citenamefont {Schweizer}\ \emph {et~al.}(2016)\citenamefont
  {Schweizer}, \citenamefont {Lohse}, \citenamefont {Citro},\ and\
  \citenamefont {Bloch}}]{Schw16}%
  \BibitemOpen
  \bibfield  {author} {\bibinfo {author} {\bibfnamefont {C.}~\bibnamefont
  {Schweizer}}, \bibinfo {author} {\bibfnamefont {M.}~\bibnamefont {Lohse}},
  \bibinfo {author} {\bibfnamefont {R.}~\bibnamefont {Citro}}, \ and\ \bibinfo
  {author} {\bibfnamefont {I.}~\bibnamefont {Bloch}},\ }\href {\doibase
  10.1103/PhysRevLett.117.170405} {\bibfield  {journal} {\bibinfo  {journal}
  {Phys. Rev. Lett.}\ }\textbf {\bibinfo {volume} {117}},\ \bibinfo {pages}
  {170405} (\bibinfo {year} {2016})}\BibitemShut {NoStop}%
\bibitem [{\citenamefont {Shindou}(2005)}]{Shin5}%
  \BibitemOpen
  \bibfield  {author} {\bibinfo {author} {\bibfnamefont {R.}~\bibnamefont
  {Shindou}},\ }\href {\doibase 10.1143/JPSJ.74.1214} {\bibfield  {journal}
  {\bibinfo  {journal} {Journal of the Physical Society of Japan}\ }\textbf
  {\bibinfo {volume} {74}},\ \bibinfo {pages} {1214} (\bibinfo {year}
  {2005})},\ \Eprint
  {http://arxiv.org/abs/https://doi.org/10.1143/JPSJ.74.1214}
  {https://doi.org/10.1143/JPSJ.74.1214} \BibitemShut {NoStop}%
\bibitem [{\citenamefont {Meidan}\ \emph {et~al.}(2011)\citenamefont {Meidan},
  \citenamefont {Micklitz},\ and\ \citenamefont {Brouwer}}]{Meid11}%
  \BibitemOpen
  \bibfield  {author} {\bibinfo {author} {\bibfnamefont {D.}~\bibnamefont
  {Meidan}}, \bibinfo {author} {\bibfnamefont {T.}~\bibnamefont {Micklitz}}, \
  and\ \bibinfo {author} {\bibfnamefont {P.~W.}\ \bibnamefont {Brouwer}},\
  }\href {\doibase 10.1103/PhysRevB.84.075325} {\bibfield  {journal} {\bibinfo
  {journal} {Phys. Rev. B}\ }\textbf {\bibinfo {volume} {84}},\ \bibinfo
  {pages} {075325} (\bibinfo {year} {2011})}\BibitemShut {NoStop}%
\bibitem [{\citenamefont {Zhou}\ \emph {et~al.}(2014)\citenamefont {Zhou},
  \citenamefont {Zhang}, \citenamefont {Sheng}, \citenamefont {Shen},
  \citenamefont {Sheng},\ and\ \citenamefont {Xing}}]{Zhou14}%
  \BibitemOpen
  \bibfield  {author} {\bibinfo {author} {\bibfnamefont {C.~Q.}\ \bibnamefont
  {Zhou}}, \bibinfo {author} {\bibfnamefont {Y.~F.}\ \bibnamefont {Zhang}},
  \bibinfo {author} {\bibfnamefont {L.}~\bibnamefont {Sheng}}, \bibinfo
  {author} {\bibfnamefont {R.}~\bibnamefont {Shen}}, \bibinfo {author}
  {\bibfnamefont {D.~N.}\ \bibnamefont {Sheng}}, \ and\ \bibinfo {author}
  {\bibfnamefont {D.~Y.}\ \bibnamefont {Xing}},\ }\href {\doibase
  10.1103/PhysRevB.90.085133} {\bibfield  {journal} {\bibinfo  {journal} {Phys.
  Rev. B}\ }\textbf {\bibinfo {volume} {90}},\ \bibinfo {pages} {085133}
  (\bibinfo {year} {2014})}\BibitemShut {NoStop}%
\bibitem [{\citenamefont {Chen}\ \emph {et~al.}(2020)\citenamefont {Chen},
  \citenamefont {Cai},\ and\ \citenamefont {Zhang}}]{Chen20}%
  \BibitemOpen
  \bibfield  {author} {\bibinfo {author} {\bibfnamefont {Q.}~\bibnamefont
  {Chen}}, \bibinfo {author} {\bibfnamefont {J.}~\bibnamefont {Cai}}, \ and\
  \bibinfo {author} {\bibfnamefont {S.}~\bibnamefont {Zhang}},\ }\href
  {\doibase 10.1103/PhysRevA.101.043614} {\bibfield  {journal} {\bibinfo
  {journal} {Phys. Rev. A}\ }\textbf {\bibinfo {volume} {101}},\ \bibinfo
  {pages} {043614} (\bibinfo {year} {2020})}\BibitemShut {NoStop}%
\bibitem [{\citenamefont {Julià-Farré}\ \emph {et~al.}(2024)\citenamefont
  {Julià-Farré}, \citenamefont {Argüello-Luengo}, \citenamefont {Henriet},\
  and\ \citenamefont {Dauphin}}]{Farre24}%
  \BibitemOpen
  \bibfield  {author} {\bibinfo {author} {\bibfnamefont {S.}~\bibnamefont
  {Julià-Farré}}, \bibinfo {author} {\bibfnamefont {J.}~\bibnamefont
  {Argüello-Luengo}}, \bibinfo {author} {\bibfnamefont {L.}~\bibnamefont
  {Henriet}}, \ and\ \bibinfo {author} {\bibfnamefont {A.}~\bibnamefont
  {Dauphin}},\ }\href {\doibase https://doi.org/10.48550/arXiv.2402.09311}
  {\enquote {\bibinfo {title} {Quantized {Thouless} pumps protected by
  interactions in dimerized rydberg tweezer arrays},}\ } (\bibinfo {year}
  {2024}),\ \Eprint {http://arxiv.org/abs/2402.09311} {arXiv:2402.09311
  [cond-mat.quant-gas]} \BibitemShut {NoStop}%
\bibitem [{\citenamefont {M\'arquez}\ \emph {et~al.}(2024)\citenamefont
  {M\'arquez}, \citenamefont {Aucar~Boidi}, \citenamefont {Hallberg},\ and\
  \citenamefont {Aligia}}]{Marquez24}%
  \BibitemOpen
  \bibfield  {author} {\bibinfo {author} {\bibfnamefont {B.~F.}\ \bibnamefont
  {M\'arquez}}, \bibinfo {author} {\bibfnamefont {N.}~\bibnamefont
  {Aucar~Boidi}}, \bibinfo {author} {\bibfnamefont {K.}~\bibnamefont
  {Hallberg}}, \ and\ \bibinfo {author} {\bibfnamefont {A.~A.}\ \bibnamefont
  {Aligia}},\ }\href {\doibase 10.1103/PhysRevB.109.235143} {\bibfield
  {journal} {\bibinfo  {journal} {Phys. Rev. B}\ }\textbf {\bibinfo {volume}
  {109}},\ \bibinfo {pages} {235143} (\bibinfo {year} {2024})}\BibitemShut
  {NoStop}%
\bibitem [{\citenamefont {Kordon}\ \emph {et~al.}(2024)\citenamefont {Kordon},
  \citenamefont {Fern\'andez},\ and\ \citenamefont {Roura-Bas}}]{Kordon24}%
  \BibitemOpen
  \bibfield  {author} {\bibinfo {author} {\bibfnamefont {F.}~\bibnamefont
  {Kordon}}, \bibinfo {author} {\bibfnamefont {J.}~\bibnamefont {Fern\'andez}},
  \ and\ \bibinfo {author} {\bibfnamefont {P.}~\bibnamefont {Roura-Bas}},\
  }\href {\doibase 10.1103/PhysRevB.110.075121} {\bibfield  {journal} {\bibinfo
   {journal} {Phys. Rev. B}\ }\textbf {\bibinfo {volume} {110}},\ \bibinfo
  {pages} {075121} (\bibinfo {year} {2024})}\BibitemShut {NoStop}%
\end{thebibliography}%

\end{document}